\documentclass[a4paper,11pt]{article}
\usepackage{pos}
\usepackage{wrapfig}
\graphicspath{{./Images/}}
\usepackage{setspace}
\usepackage{natbib}
\title{Photon-jet correlations in p-p and Pb-Pb collisions using JETSCAPE framework}
\ShortTitle{$\gamma$-jet correlations using JETSCAPE framework}

\author*[1]{Chathuranga Sirimanna}

\affiliation{Department of Physics and Astronomy, Wayne State University, Detroit, Michigan 48201, USA}

\note{For the JETSCAPE collaboration.}

\emailAdd{chathuranga.sirimanna@wayne.edu}

\abstract{It is now well established that jet modification is a multistage effect; hence a single model alone cannot describe all facets of jet modification. The JETSCAPE framework is a multistage framework that uses several modules to simulate different stages of jet propagation through the QGP medium. These simulations require a set of parameters to ensure a smooth transition between stages. We fine tune these parameters to successfully describe a variety of observables, such as the nuclear modification factors of leading hadrons and jets, jet shape, and jet fragmentation function.

Photons can be produced in the hard scattering or as radiation from quarks inside jets. In this work, we study photon-jet transverse momentum imbalance and azimuthal correlation for both $p-p$ and $Pb-Pb$ collision systems. All the photons produced in each event, including the photons from hard scattering, radiation from the parton shower, and radiation from hadronization are considered with an isolation cut to directly compare with experimental data. The simulations are conducted using the same set of tuned parameters as used for the jet analysis. No new parameters are introduced or tuned. We demonstrate a significantly improved agreement with photons from $Pb-Pb$ collisions compared to prior efforts. This work provides an independent, parameter free verification of the multistage evolution framework.}

\FullConference{%
  HardProbes2020\\
  1-6 June 2020\\
  Austin, Texas}

\begin{document}
\maketitle

\section{Introduction}

Prompt photons are produced directly in hard sub-processes, and most of the time they are produced even before the production of quark-gluon plasma (QGP) in relativistic heavy-ion collisions. They are often identified as a calibrated probe of QGP since they can be used to estimate the energy and the direction of jet-initiating partons in photon-jet events before energy loss. Even though identifying prompt photons is relatively easy in theoretical simulations, it is impossible to exactly pinpoint them in actual high energy particle collisions. Hence, an isolation criterion is necessary to identify them. The isolation criteria are different for different experiments depending on the detector specifications and methods of analysis \cite{CMS:2013oua, Aaboud:2018anc}. Since photons do not interact with the QGP medium, usually they are isolated within a cone of a significant radius. But there can be isolated photons within the criteria that are not prompt photons. Even though that is a very small portion of all isolated photons, that can have some effect on the final results.

In this study two of the most widely used observables related to photon triggered jets, photon-jet transverse momentum imbalance ($x_J$ distribution) and azimuthal correlations ($\Delta \phi$ distribution), are  studied using JETSCAPE framework \cite{Cao:2017zih, Putschke:2019yrg}. We followed the experimental analysis criteria exactly in order to make a better comparison with experimental results. Since there is no fine-tuning on either the framework or any modules used inside the framework, this can be considered as an independent, parameter-free verification of the multistage evolution framework.

\section{Photons in the JETSCAPE framework}

JETSCAPE is a general, modular, and extensive framework that can be used to simulate both p-p and heavy-ion collisions. Since no single model can describe all stages of jet evolution simultaneously, JETSCAPE uses a multistage approach. In this multistage approach, the JETSCAPE framework switches between different stages depending on the virtuality and the energy of the intermediate shower partons. There are number of different modules available in the framework by default. In this study, PYTHIA is used for the initial hard scattering. PYTHIA is terminated after the hard scattering and all the partons are fed into the energy loss modules. The intermediate shower is done inside the energy loss modules (in this case only MATTER \cite{Majumder:2013re, Majumder:2009zu} for p-p and MATTER combined with LBT \cite{Wang:2013cia, PhysRevC.91.054908} for Pb-Pb). After the intermediate shower, all the partons were again fed into Pytha for string hadronization using the colorless hadronization scheme developed inside the framework. In the colorless hadronization, the color information is randomly assigned to the partons before hadronizing them using PYTHIA string fragmentation. Pre-generated hydro profiles with initial conditions were used for Pb-Pb collisions. The leading hadron and jet results for both p-p \cite{Kumar:2019bvr} and Pb-Pb \cite{Park:2019sdn, Kumar:2020vkx} collisions are available in the literature. The same set of parameters were used in this study without any further tuning. 

Even though the medium-induced terms for energy loss are included in the energy loss modules, medium-induced photon emission terms are not included. Therefore there is more room for improvement in theoretical approaches. All the photons from the shower are included in this analysis. That includes the prompt photons, intermediate shower photons, and the photons radiated by hadrons in the process of hadronization. Isolated photons were identified by using the same isolation criteria and the same azimuthal cut used by the experimental analysis. The azimuthal cut is used to make sure that the photon and the leading jet are back to back. A cone surrounding the direction of a photon is used to describe an isolated photon. The radius of the cone, $\Delta R$ is identical to the experiment (CMS and ATLAS in this case). If the total transverse momentum, $p_T$ within the cone is less than a given isolation momentum, $p_T^{iso}$, the photon is considered to be an isolated photon. Since photon triggered jets are rare to observe, 27 million events (in 54 pT hat bins) were used for the p-p analysis at $2.76~TeV$ and 66 million events (in 66 pT hat bins) were used for the Pb-Pb analysis at $5.02~TeV$.

\section{Results and discussion}

Both $x_J$ and $\Delta \phi$ distributions are studied for p-p collisions at 2.76 TeV. The Fastjet software package is used to cluster jets using the anti-$k_T$ jet clustering algorithm with jet radius, $R_{jet} = 0.3$. The parameters are set to $\Delta R = 0.4$ and $p_T < 5 GeV$ to idenitfy the isolated photons. The transverse momentum of the jet, $p_T^{jet}$, and the rapidity, $\eta$, interval are set to $p_T^{jet} > 30~GeV$, $\left| \eta_\gamma \right| > 1.44$, and $\left| \eta_{jet} \right| > 1.6$ to make a reasonable comparison with the experimental results. The A-14 tune is used to generate PYTHIA simulations as it is the only tune to give better description of experimental results. Since the experimental results are smeared to make a better comparison with Pb-Pb results, the same smearing function \cite{CMS:2013oua} is used to generate both JETSCAPE and PYTHIA results.

\begin{figure} [ht]
    \addtolength{\abovecaptionskip}{-3mm}
	\begin{minipage} [l] {0.3\textwidth}
	\centering
	\includegraphics[width=\textwidth]{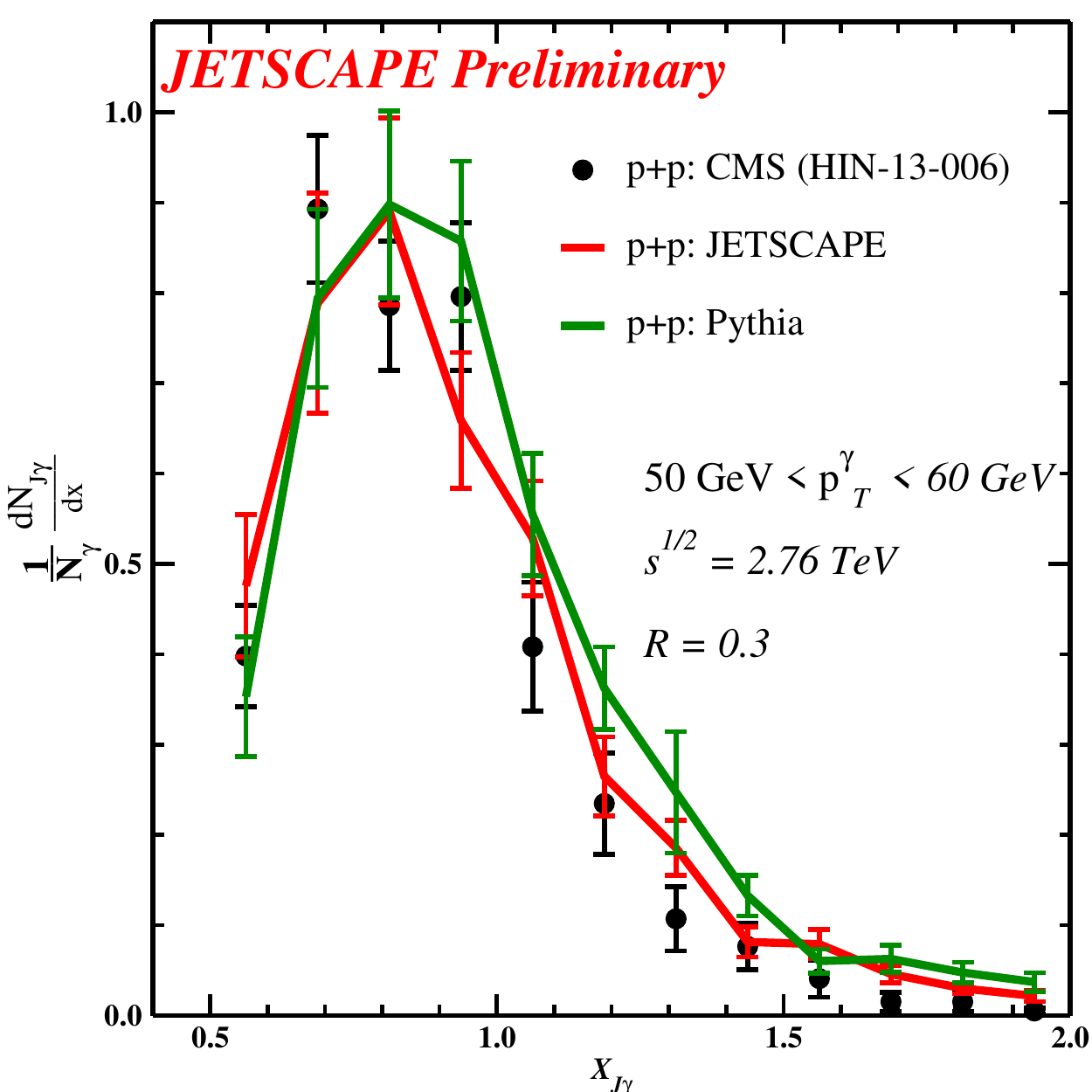}
	\end{minipage}
	\hspace{0.5cm}
	\begin{minipage} [r] {0.3\textwidth}
	\centering
	\includegraphics[width=\textwidth]{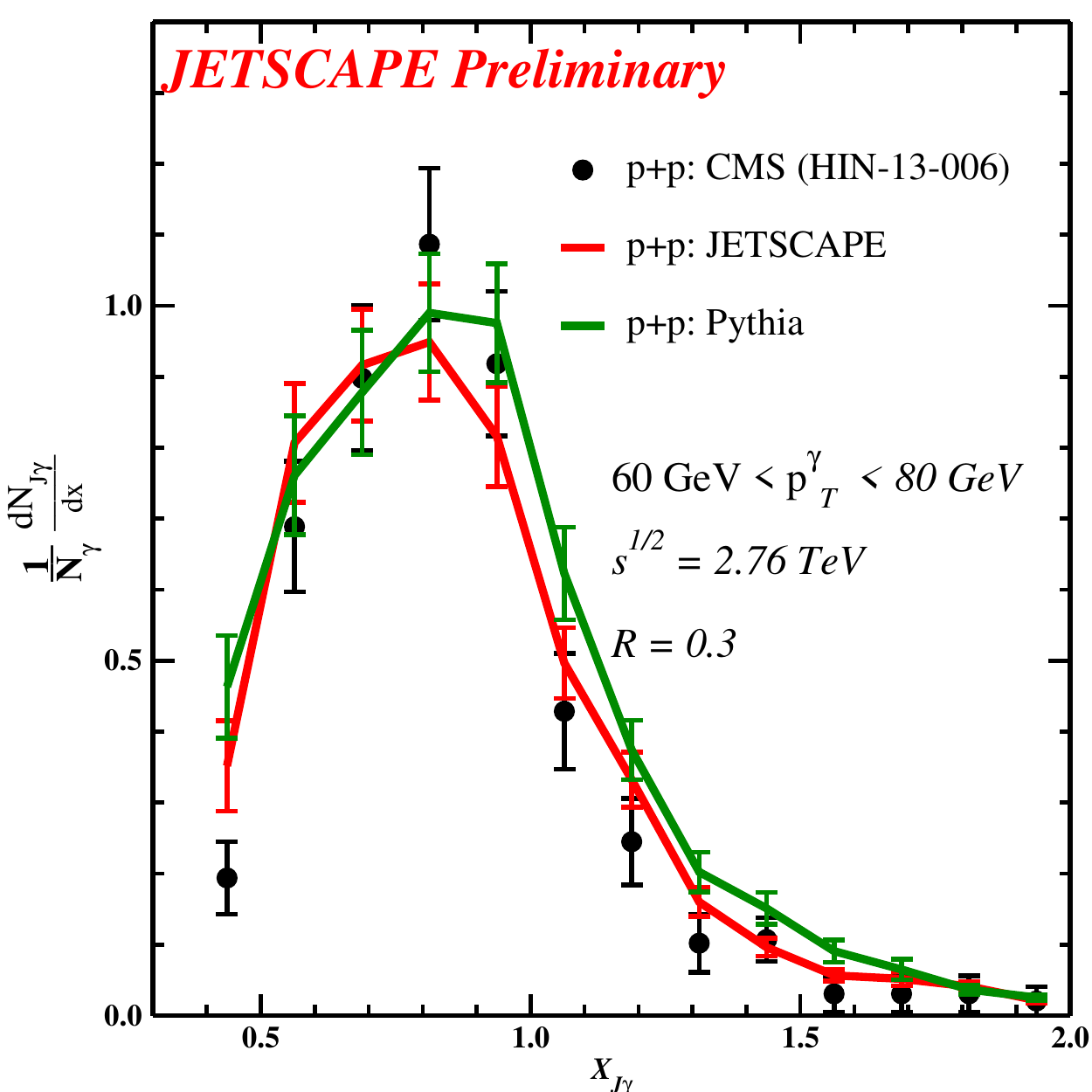}
	\end{minipage}
	\hspace{0.5cm}
	\begin{minipage} [r] {0.3\textwidth}
	\centering
	\includegraphics[width=\textwidth]{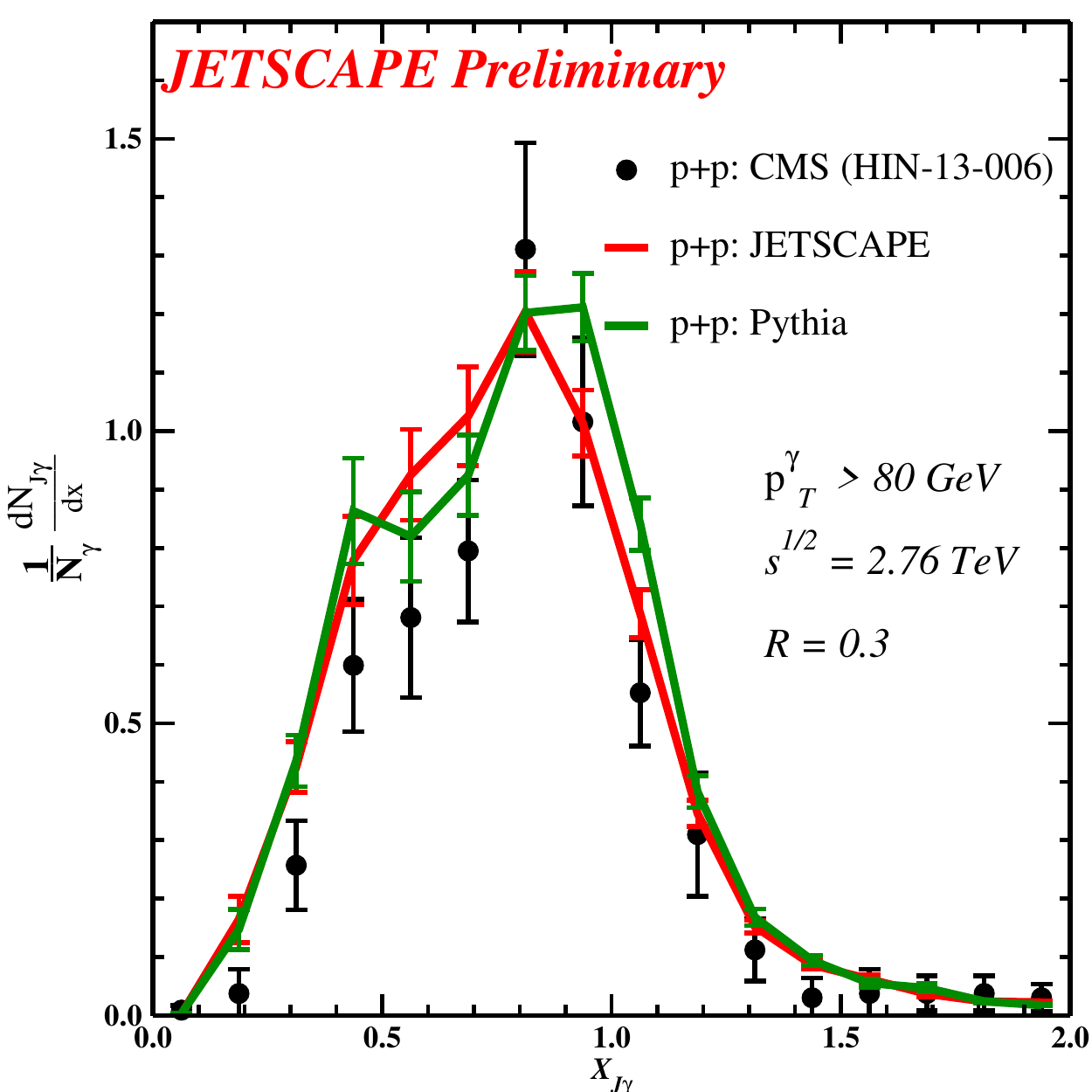}
	\end{minipage}
	\caption{Photon-Jet transverse momenum imbalance results for p-p collisions from JETSCAPE and PYTHIA(A14 tune) compared with CMS results.}
	\label{fig:XJ_PP_2760CMS}
\end{figure}

Figure \ref{fig:XJ_PP_2760CMS} shows the $x_J$ distributions for three different $p_T^\gamma$ intervals. An azimuthal angle between the leading jet and the photon was restricted by using the condition $\left| \Delta \phi_{ \gamma-jet} \right| > \frac{ 7 \phi }{ 8 }$ in order to make sure the photon and the leading jet are indeed back to back. As seen in Figure \ref{fig:XJ_PP_2760CMS}, both JETSCAPE and PYTHIA provide excellent descriptions of the experimental results. 

\begin{figure}[ht]
    \addtolength{\abovecaptionskip}{-3mm}
	\begin{minipage} [l] {0.45\textwidth}
	\centering
	\includegraphics[width=0.8\textwidth]{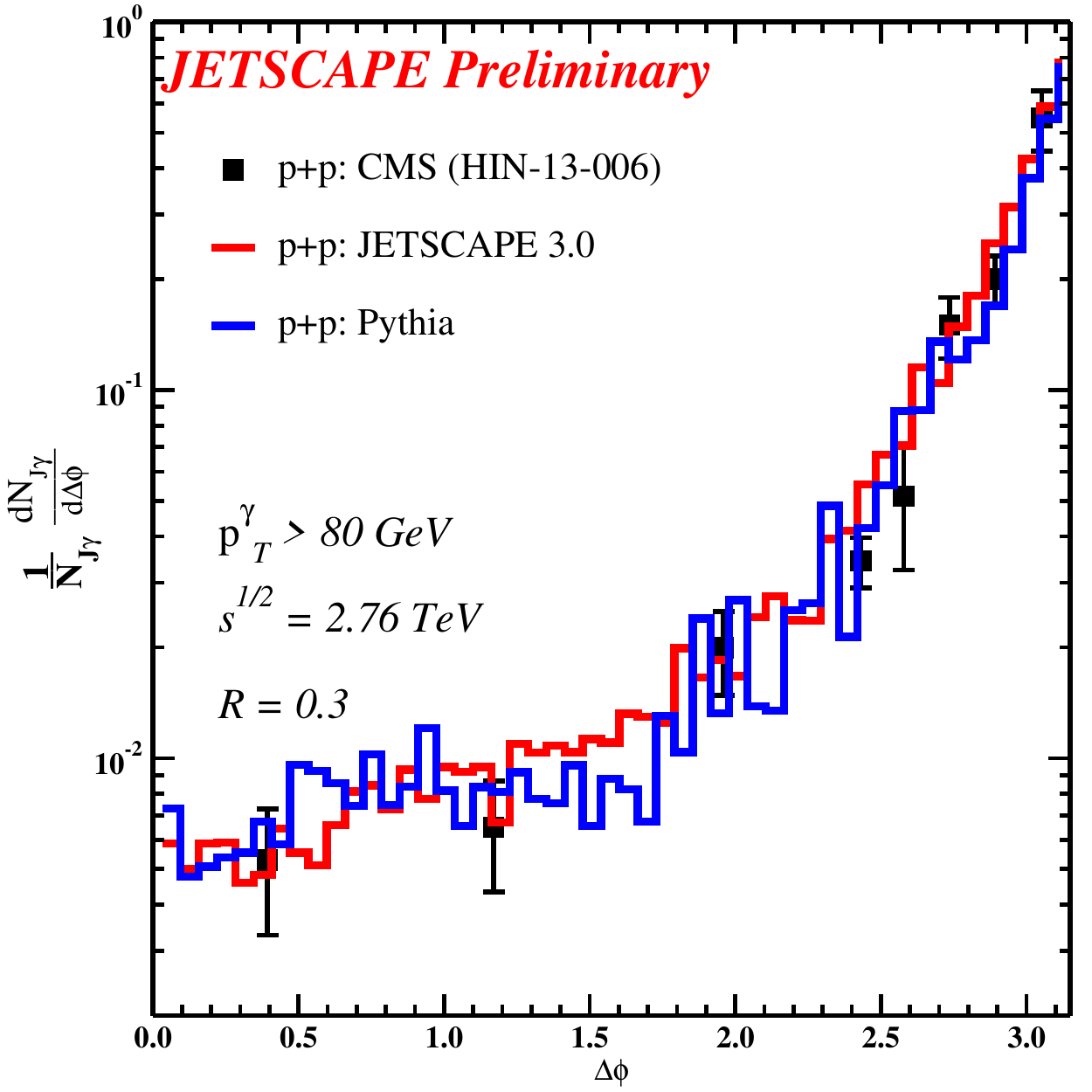}
	\end{minipage}
	\hspace{0.5cm}
	\begin{minipage} [r] {0.45\textwidth}
	\centering
	\includegraphics[width=0.8\textwidth]{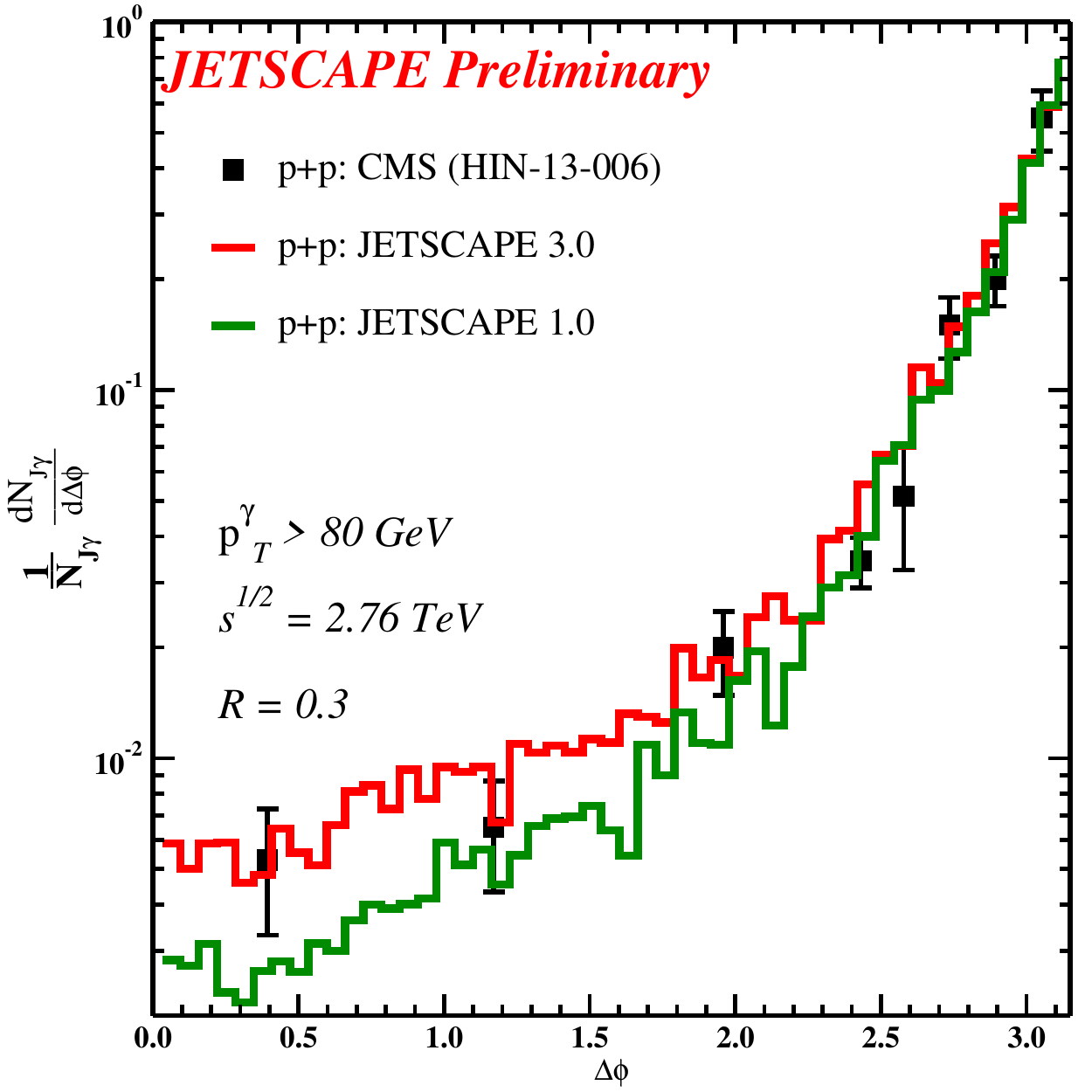}
	\end{minipage}
	\caption{Photon-Jet azimuthal correlation results for p-p collisions at $2.76~TeV$ from JETSCAPE (1.0 and 3.0) and PYTHIA(A14 tune) compared with CMS results.}
	\label{fig:Dphi_PP_2760CMS}
\end{figure}

Figure \ref{fig:Dphi_PP_2760CMS} shows the $\Delta \phi$ distributions  for $p_T^\gamma > 80 GeV$. One plot compares the experimental results with JETSCAPE and PYTHIA results, and the other plot compares the experimental results with two versions of JETSCAPE, 1.0 and 3.0. As seen in Figure \ref{fig:Dphi_PP_2760CMS}, excellent agreement is observed between experimental results and the theoretical results by JETSCAPE 3.0 and PYTHIA.  Since JETSCAPE 1.0 did not support intermediate shower photons, we see a significant suppression in the low $\Delta \phi$ region.

\begin{figure}[ht]
    \addtolength{\abovecaptionskip}{-3mm}
	\begin{minipage} [l] {0.45\textwidth}
	\centering
	\includegraphics[width=0.8\textwidth]{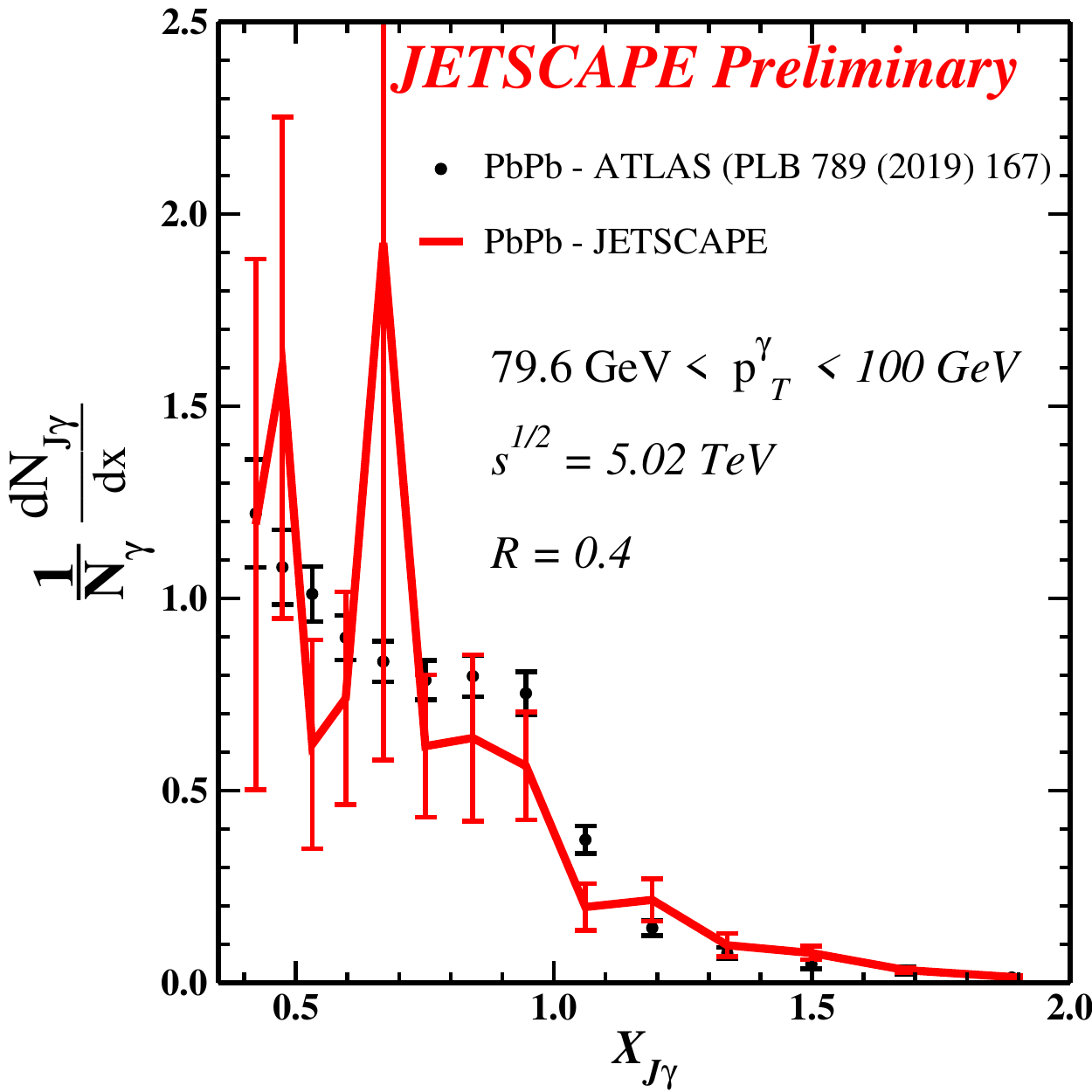}
	\end{minipage}
	\hspace{0.5cm}
	\begin{minipage} [r] {0.45\textwidth}
	\centering
	\includegraphics[width=0.8\textwidth]{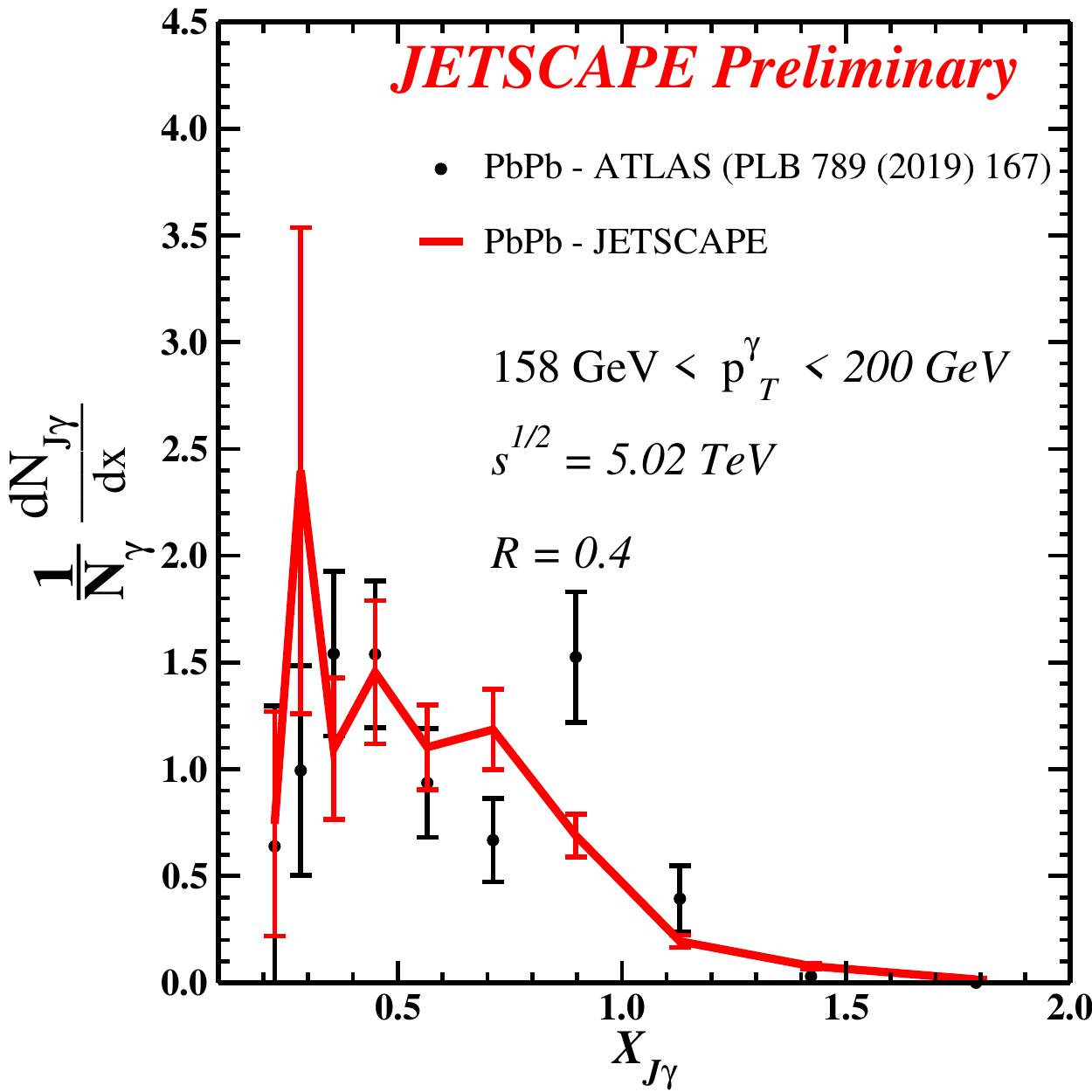}
	\end{minipage}
	\caption{Photon-Jet transverse momenum imbalance results for Pb-Pb collisions at $5.02~TeV$ from JETSCAPE compared with ATLAS results.}
	\label{fig:XJ_PbPb_5020ATLAS}
\end{figure}

The $x_J$ distribution generated for most central $(0-10\%)$ Pb-Pb collisions at $5.02~TeV$ is shown in Figure \ref{fig:XJ_PbPb_5020ATLAS}. The anti-$k_T$ jet clustering algorithm with jet radius, $R_{jet} = 0.4$ is used to cluster jets through Fastjet. The parameters are set to $\Delta R = 0.3$ and $p_T < 8 GeV$ to identify the isolated photons. The cutoff values for $p_T^{jet}$ and $\eta$ interval are set to $p_T^{jet} > 31.6~GeV$, $\left| \eta_\gamma \right| > 2.37$ with exclusion of $1.37<\left| \eta_\gamma \right|<1.52$ region, and $\left| \eta_{jet} \right| > 2.8$. Since the amount of data generated is statistically insufficient to describe the low $x_J$ region of the spectrum, we are currently generating more statistics to obtain better results for these observables.

\section{Summary and outlook}

Both Photon-jet transverse momentum imbalance and azimuthal correlation results from JETSCAPE are in excellent agreement with experimental results from CMS in p-p collisions at $2.76~TeV$. Also, JETSCAPE can describe Photon-jet transverse momentum imbalance results from ATLAS in Pb-Pb collisions at $5.02~TeV$ with some statistical fluctuations at low $x_J$ region. This demonstrates that the multi-stage evolution can describe all the stages of jet evolution significantly better than single module evolution. Also the same set of parameters used to describe leading-hadron and jet spectrum are used to describe photon observables. This emphasize that the JETSCAPE framework can describe most observables using the same set of parameters for different center of mass energies. Therefore, this can be considered as a parameter-free verification of multi-stage evolution.

\acknowledgments

These proceedings are supported in part by the National Science Foundation (NSF) within the framework of the JETSCAPE collaboration, under grant numbers ACI-1550300 and in part by the U.S. Department of Energy (DOE) under grant number DE-SC0013460.


\bibliographystyle{JHEP} 
\footnotesize\bibliography{Sirimanna_C}


\end{document}